\documentclass[elsart12,epsf,eqsecnum]{revtex4}

\usepackage{epsfig,epsf}
\begin{document}

%%%%%%%%%%%%%%%%%%%%%%%%%%%%%%%%%%%%%%%%%%%%%%%%%%%%%%%%%%%%%%%%%%%
%\documentclass[preprint,paper]{JHEP3} % 10pt is ignored!

%\JHEP{00(2002)000}

%\JHEPspecialurl{http://jhep.sissa.it/JOURNAL/JHEP3.tar.gz}

%\usepackage{epsfig,multicol}
%\usepackage{latexsym,amsmath,amssymb,bm}

%%%%%%%%%%%%%%%%%%%%%%%% latex 2 eps %%%%%%%%%%%%%%%%%%%%%%%%%%

%\documentstyle[epsf,epsfig,aps]{revtex}

%\documentclass[12pt,prl,aps]{revtex4}

%\documentclass[12pt]{article}

%\usepackage{epsfig,epsf}

%\usepackage{showkeys}
%\usepackage{cite}

%\voffset1.5cm
\def\beq{\begin{equation}}

\def\eeq{\end{equation}}
\def\eq#1{{Eq.~(\ref{#1})}}

\def\fig#1{{Fig.~\ref{#1}}}
             
\newcommand{\bas}{\bar{\alpha}_S}

\newcommand{\as}{\alpha_S} 

\newcommand{\bra}[1]{\langle #1 |}

\newcommand{\ket}[1]{|#1\rangle}

\newcommand{\bracket}[2]{\langle #1|#2\rangle}

\newcommand{\intp}[1]{\int \frac{d^4 #1}{(2\pi)^4}}

\newcommand{\mn}{{\mu\nu}}

\newcommand{\tr}{{\rm tr}}

\newcommand{\Tr}{{\rm Tr}}

\newcommand{\T} {\mbox{T}}
\newcommand{\W} {\text{W}}

\newcommand{\braket}[2]{\langle #1|#2\rangle}

\newcommand{\ab}{\bar{\alpha}_S}

\newcommand{\bea}{\begin{eqnarray}}

\newcommand{\eea}{\end{eqnarray}}

\setcounter{secnumdepth}{7}

\setcounter{tocdepth}{7}

\parskip=\itemsep               %?

%\setlength{\itemsep}{0pt}       %?

%\setlength{\partopsep}{0pt}     %?

%\setlength{\topsep}{0pt}        %?

%---layout fuer eine dina4 seite-------------------

%\setlength{\textheight}{22cm}

\setlength{\textwidth}{168mm}

%\setlength{\topmargin}{.5cm}

%\input psfig

%%%%%%%%%%%%%%%%%%%%%%%%%%%%%%%%%%%%%%%%%%%%%%%%%%%%%%%%%%%%%%%

%\renewcommand{\thefootnote}{\fnsymbol{footnote}}

\newcommand{\beqar}[1]{\begin{eqnarray}\label{#1}}

\newcommand{\eeqar}{\end{eqnarray}}

\newcommand{\m}{\marginpar{*}}

\newcommand{\lash}[1]{\not\! #1 \,}

\newcommand{\nn}{\nonumber}

\newcommand{\D}{\partial}

\newcommand{\h}{\frac{1}{2}}

\newcommand{\g}{{\rm g}}

\newcommand{\el}{{\cal L}}

\newcommand{\A}{{\cal A}}

\newcommand{\Ka}{{\cal K}}

\newcommand{\al}{\alpha}

\newcommand{\be}{\beta}

\newcommand{\ep}{\varepsilon}

\newcommand{\ga}{\gamma}

\newcommand{\de}{\delta}

\newcommand{\De}{\Delta}

\newcommand{\et}{\eta}

\newcommand{\ka}{\vec{\kappa}}

\newcommand{\la}{\lambda}

\newcommand{\ph}{\varphi}

\newcommand{\si}{\sigma}

\newcommand{\ro}{\varrho}

\newcommand{\Ga}{\Gamma} 

\newcommand{\om}{\omega}

\newcommand{\La}{\Lambda}  

\newcommand{\tG}{\tilde{G}}

\renewcommand{\theequation}{\thesection.\arabic{equation}}

%%%%%%%%%%%%%%%%%%%%%%%%%%%%%%%%%%%%%%%%%%%%%%%%%%%%%%%%%%%%%%%%%

% ABBREVIATED JOURNAL NAMES  

%

\def\ap#1#2#3{     {\it Ann. Phys. (NY) }{\bf #1} (19#2) #3}

\def\arnps#1#2#3{  {\it Ann. Rev. Nucl. Part. Sci. }{\bf #1} (19#2) #3}

\def\npb#1#2#3{    {\it Nucl. Phys. }{\bf B#1} (19#2) #3}

\def\plb#1#2#3{    {\it Phys. Lett. }{\bf B#1} (19#2) #3}

\def\prd#1#2#3{    {\it Phys. Rev. }{\bf D#1} (19#2) #3}

\def\prep#1#2#3{   {\it Phys. Rep. }{\bf #1} (19#2) #3}

\def\prl#1#2#3{    {\it Phys. Rev. Lett. }{\bf #1} (19#2) #3}

\def\ptp#1#2#3{    {\it Prog. Theor. Phys. }{\bf #1} (19#2) #3}

\def\rmp#1#2#3{    {\it Rev. Mod. Phys. }{\bf #1} (19#2) #3}

\def\zpc#1#2#3{    {\it Z. Phys. }{\bf C#1} (19#2) #3}

\def\mpla#1#2#3{   {\it Mod. Phys. Lett. }{\bf A#1} (19#2) #3}

\def\nc#1#2#3{     {\it Nuovo Cim. }{\bf #1} (19#2) #3}

\def\yf#1#2#3{     {\it Yad. Fiz. }{\bf #1} (19#2) #3}

\def\sjnp#1#2#3{   {\it Sov. J. Nucl. Phys. }{\bf #1} (19#2) #3}

\def\jetp#1#2#3{   {\it Sov. Phys. }{JETP }{\bf #1} (19#2) #3}

\def\jetpl#1#2#3{  {\it JETP Lett. }{\bf #1} (19#2) #3}

%%%%%%%%% notice the parenthesys is only on one side

\def\ppsjnp#1#2#3{ {\it (Sov. J. Nucl. Phys. }{\bf #1} (19#2) #3}

\def\ppjetp#1#2#3{ {\it (Sov. Phys. JETP }{\bf #1} (19#2) #3}

\def\ppjetpl#1#2#3{{\it (JETP Lett. }{\bf #1} (19#2) #3} 

\def\zetf#1#2#3{   {\it Zh. ETF }{\bf #1}(19#2) #3}

\def\cmp#1#2#3{    {\it Comm. Math. Phys. }{\bf #1} (19#2) #3}

\def\cpc#1#2#3{    {\it Comp. Phys. Commun. }{\bf #1} (19#2) #3}

\def\dis#1#2{      {\it Dissertation, }{\sf #1 } 19#2}

\def\dip#1#2#3{    {\it Diplomarbeit, }{\sf #1 #2} 19#3 }

\def\ib#1#2#3{     {\it ibid. }{\bf #1} (19#2) #3}

\def\jpg#1#2#3{        {\it J. Phys}. {\bf G#1}#2#3}  

%

%%%%%%%%%%%%%%%%%%%%%%%%%%%%%%%%%%%%%%%%%%%%%%%%%%%%%%%%%%%%%%%%%%%%%

%

%\renewcommand{\thefigure}{{\protect\bf\arabic{figure}}}

\def\thefootnote{\fnsymbol{footnote}}

\title{
 Universal hydrodynamics and charged hadron multiplicity at the LHC }
\author{M. Lublinsky$^a$ and E. Shuryak$^b$   \\ }
\address{$^a$Physics Department, Ben Gurion University of the Negev, Beer Sheva 84105, Israel \\
$^b$ Physics Department, SUNY Stony Brook,  NY 11794-3800, USA}

\begin{abstract}
Time evolution of a "little bang" created in heavy ion collisions can be divided into two phases, the pre-equilibrium and hydrodynamic.  
At what moment the evolution becomes  hydrodynamic and is there any universality in the hydrodynamic flow? To answer these questions 
we briefly discuss various versions of hydrodynamics and their applicability conditions. In particular, we elaborate on 
the idea of ``universal" (all-order resumed) hydrodynamics and propose a simple new model for it. The model is motivated by  results obtained recently 
via the AdS/CFT correspondence.  Finally, charged hadron multiplicities in heavy ion collisions at the RHIC and  LHC are discussed. 
At the freezout, the multiplicities can be related to  total entropy produced in the collision.
Assuming the universal hydrodynamics to hold, we calculate the  entropy production in the hydro stage of the collision.
We end up speculating about a connection between the multiplicity growth and the temperature dependence of the QGP viscosity.    
\end{abstract}
\maketitle
%\keywords{qcd,dis}

%\renewcommand{\thefootnote}{\fnsymbol{footnote}}

%\begin{document}

%*********************************************************************************  

%*********************************************************************************  

%\def\thefootnote{\arabic{footnote}} 

\section{Introduction}

 This paper contains some further developments of the ideas put forward in our paper \cite{LS1}.  There we argued  that 
 entropy production in the strongly coupled quark gluon plasma (sQGP) should be computed using an all-order resummed hydrodynamics and  that the resummation makes it possible
 to provide reliable estimates even starting from very short thermalization 
times. The main goal of this note  is to connect this proposal to some recent theory developments based on the AdS/CFT setting \cite{Malda}, which  support our ideas,
as well as to address the phenomenological question of charged  particle multiplicity production in heavy ion collisions at the LHC,
to be detailed below in section \ref{sec_mult}. Let us stress here that the entropy production 
is only one of several  applications, for which an all-order resummation might be important. There are additional 
interesting phenomena, in which  matter gradients are large and applicability limits of standard hydrodynamics is in question. Let us give here two examples of those.

 As recent studies have shown,  fluctuations of  initial state
density in heavy ion collisions are the origin of sound waves.  By freezeout, these waves reach
 large distances,  comparable to the fireball radius itself, and are observed as fluctuations of  angular harmonics in the 
particle distributions. It is remarkable that amplitudes of up to  9-th harmonics  have been measured, displaying good agreement with
hydrodynamics \cite{Lacey:2011av,Staig:2011wj,Staig}. Yet, the questions how to treat  these fluctuations in non-equilibrium and  from what initial times can they 
 be evolved hydrodynamically   remains unanswered.

 ``Mach cones" induced in the matter by quenching jets \cite{Casalderrey-Solana:2004qm,Horst} present another application of the sound waves in heavy ion physics. Unlike sounds from the previous example, 
the jet-induced waves were studied in detail within the AdS/CFT context \cite{CY}. The results were shown to have a  good agreement with hydrodynamics at later stages but
 when exactly hydro becomes applicable and why  still could have been studied more, given the
 exact AdS/CFT solution. The issue becomes even  more important 
with the first LHC data on jets, revealing events with huge amounts of energy, $\sim 100\, GeV$,  deposited by a jet.
This calls for studies of the full nonlinear settings, beyond the linearized sound wave approximation. 
  
In section III, we  discuss   initial conditions for hydrodynamics from the perspective of the AdS/CFT results. We also propose in this section a new,
all-order resumed, hydrodynamics  model for Bjorken explosion. In Section IV, we use this model in order to compute the entropy production in the hydro phase.
Phenomenological relevance  to the data on charged particle multiplicities  is also discussed. We summarize and provide additional discussions  in section V.

\section{Multiplicities in $pp$ and $AA$ collisions} \label{sec_mult}
One of the first discoveries made by the LHC  is a rapid rise with energy of  multiplicities of charged hadrons produced both in pp
and heavy ion collisions. The discovery is especially dramatic in heavy ions collisions, where most of the existing models have failed to predict the data.

The first ALICE data  on charged particle multiplicity in  lead lead collisions are
\cite{ALICE}:
\beq {dN\over d\eta}|_{PbPb}(2.76 TeV)\,=\,1584\,\pm76, \eeq
 combined with the earlier data from the RHIC, these ones imply  the multiplicity in AA collisions growing with the (center of mass) energy per nucleon as
 \beq {dN\over d\eta}|_{PbPb}(E_{NN})\sim E_{NN}^{0.30}\,. \eeq
The corresponding power in the $pp$ collisions is $0.22$, and thus the ratio of the two also grows with the energy
 \beq {dN\over d\eta}|_{PbPb}\,\,/\,\, {dN\over d\eta}|_{pp} \sim  E_{NN}^{0.08} \,.\eeq
 From the RHIC energy ($E=0.2\, TeV$) to the LHC, the double ratio is 
  \beq \label{dr}
  {{dN\over d\eta}|_{PbPb,LHC}\,\,/\,\, {dN\over d\eta}|_{pp,LHC} \over  {dN\over d\eta}|_{AuAu,RHIC}\,\,/\,\, {dN\over d\eta}|_{pp,RHIC}} = 1.23\,. \eeq
This  noticeable change with the energy  calls for a theoretical explanation. (An increase in the atomic number, 197 for Au and 208 for Pb,
explains only 0.055 of it.) 

Particle production  in heavy ion collisions proceed via two basic phases: 
(i) a pre-thermalization phase and (ii) a hydrodynamical stage. 
Theoretical frameworks used for their descriptions are very different. 

 The first one is  based on pQCD cascade of gluons, described by high energy evolution equations  including gluon saturation effects,
 or color glass condensate (CGC).  CGC relies on emergence of a semi-hard scale,
  the  saturation momentum  \beq Q_s\sim A^{1/3}\,x^{-\lambda},\ \ \ \ \ \ \lambda=0.25-0.30 \eeq related to the density of
 gluons with longitudinal momentum fraction $x$. Within the CGC approach, many quantities become universal 
 and simply scale with the saturation scale, the property known as a geometrical scaling.  As an example of this, particle's $p_t$ spectra  in $pp$ collisions
 are found to have the dependence of the type  $f(p_t/Q_s)$ \cite{McLerran:2010wm}.
 
 If the hypothesis of geometrical scaling is true, then a  CGC-based estimate for the AA/pp multiplicity ratio should be energy independent
 (see, however,  Ref.  \cite{Lappi} discussing the  DGLAP effect on  the saturation scale). 
 Yet, experiments observe a prominent growth with energy.
Another  observation is that the CGC-based multiplicity estimates tend to underestimate it at the LHC. In particular,
Ref.  \cite{Albacete}  underestimates  the observed multiplicity  by approximately 35\%:
 ${dN/ d\eta}|_{PbPb}(2.76 TeV)\simeq 1175$.
 %, the ratio $R\simeq 1.35$.
  
%It is important to stress, however, that  the present CGC-based computations are based on certain
%uncontrollable approximations, such as the $k_t$ factorization, which is approximate and can  be violated to some accuracy.

The second phase of heavy ion collision process  is hydrodynamic, and it produces particles (entropy) due to finite viscosity.
While the viscosity itself grows, from 
   strongly coupled regime   at  the beginning of the evolution to hadronic matter at its end, and even gets very large near freezeout,
  the main entropy production still happens at the very beginning. This is so because the viscosity coefficient gets multiplied by  flow gradients, which are fast decreasing with the evolution time. Below, we  will discuss  the effect of viscosity  on the multiplicity growth. 
  
%can be thought as established by now. Finite viscosity leads inevitably to additional multiplicity production, especially if fast termalization takes place. 
%Here we would like to extend our original idea of entropy production in the sQGP
%using all-order hydrodynamics \cite{LS1}, that makes it possible to provide reliable estimates even starting from very short thermalization 
%times.  The goal is to use all-order hydro to make up for the discrepancy between the numbers above, and to give some predictions for
%ALICE at 5.5 TeV.  
%$${dN\over d\eta}(5.5 TeV)\simeq 1390\ \ \ \ \ \ (CGC)$$
%
 %\cite{Albacete}
%Multiplicity rises faster in AA than in pp. 
%We assume here it is due to hydro phase which is present in AA and presumably does not in pp.

\section{When does the hydro stage start?}

This question is not well posed unless we specify what exactly is meant  by ``hydro"  and by its ``start".  To define a starting moment 
 is relatively easy:  for any theory and an approximation to it, the approximation is considered as valid as long as the two  deviate from each other 
 within a preset accuracy (say  one percent).
% it can be defined by deviations from it being smaller than some specified accuracy required from it 

The  question of defining "hydro" has different meaning and depends on what approximation is used. We will mention three cases here: \\
(i) ``ideal hydrodynamics" is a collective description that  includes local quantities only, such as  pressure and energy
density. Its accuracy/validity  depends on  viscous corrections to this local approximation, which contain  first gradients of the  flow of matter. \\
(ii) Navier-Stokes  hydrodynamics (NS) includes these viscous terms, and  its accuracy is estimated by $next$ terms involving  $two$ gradients.\\  
(iii)  ``resummed  hydrodynamics" (RH) which includes in some approximate form all higher order gradients.  Accuracy of this approximation
is given by deviations from  first principle non-equilibrium  calculations.\\
Obviously, as the accuracy of approximation increases from (i) to (iii), its applicability regions widens. In connection with
heavy ion collision processes,  it means ``the beginning of the hydro stage" moves towards  earlier and earlier times.

\subsection{Conformal ``resummed  hydrodynamics"}

When talking about all-order resummed hydro it is convenient to introduce viscosity as a momenta-dependent function. In  \cite{LS1}
we extracted it from  an AdS/CFT computed sound dispersion curve.  In \cite{LS2} we took a more formal approach, which lead us to propose the following model
\beq\label{mod1}
\eta(\omega,k^2)\,=\,{\eta_0\over 1\,-\,1/2\,k^2\,-\,i\,\omega\,\tau_R}\,.
\eeq
 Here  $\eta_0=1/2$ in dimensionless units in which $2\pi T=1$ and that corresponds to the celebrated ratio of viscosity to entropy density equal $1/4\pi$ \cite{KSS}.
  In this units, 
 $\tau_R=2-\log{2}$ and is the relaxation time of the Israel-Stewart (IS) model \cite{IS}. The model (\ref{mod1})
  reproduces well the small $\omega$ and $k$ expansion up to  fifth order. 
 
 We consider Bjorken flow \cite{Bj} as a model for the explosion. It has the simplest geometry: there is no dependence on two transverse coordinates, as well as
 on space-time rapidity $y=(1/2)\,\ln[(t-x)/(t+x)]$. What is left is a dependence on the proper time $\tau=\sqrt{t^2-x^2}$ only. In these coordinates, the metric
 is  $ds^2=-d\tau^2+\tau^2 dy^2+  d\vec{x}_{\perp}^2$, and we will not write any further details, as those are well known. 
  In the Bjorken flow, there are no spatial variations ($\vec{k}=0$) and our model (\ref{mod1}) reduces back to IS.  
It is well known that additional non-linear terms contribute to the entropy production that is not
governed  by the viscosity term only.  However, the entropy is produced mostly at the beginning of the expansion, when viscous terms are dominant.
It is especially true for the case of  very early thermalization. This is why a more or less reliable estimate of entropy production can emerge
only if we know the dissipation tensor at very large  $\omega$.

Let introduce the dimensionless variable $w=\tau\,T$.
Then, within the all order hydrodynamic approximation, the entropy production equation can be written with some ``universal function" of this variable
\beq\label{eq1}
{d w\over d\ln \tau}\,=\,F(w)\,,
\eeq
Solving (\ref{eq1}) one finds time dependence of the temperature, from  the initial time $\tau_i$ to the final (freezeout) time $\tau_f$
\beq\label{sol}
\tau(w_f)= \tau(w_i)\,exp\left[\int_{w_i}^{w_f}{dw^\prime\over F(w^\prime)}\right]   \hspace{2cm} T(w)= w/\tau(w).
\eeq
The $final$ values $T_f,\,\tau_f$ should be read off the experimental data (there are evidences that $T_f$ is about the same at the RHIC and LHC 
while $\tau_f$ grows with $E_{NN}$, and hence the total entropy (multiplicity) grows too). 

From these experimental data,  one may use the solution and
trace back to the  $initial$ values for the thermalization time and  temperature. However,  eq. (\ref{sol}) provides 
only one relation between the two. In the plane ($\tau_i\,,T_i$) it defines a curve. (This is similar to field theory RG flows of couplings).
An additional condition, to be detailed below,   is needed, in order to fix the absolute values of the initial conditions.

The function $F(w)$ can be expanded in powers of $1/w$ with coefficients of the expansion being  higher order viscosities. Thanks to the AdS/CFT  correspondence,
for conformal ${\cal N}=4$ plasma
the expansion terms are known up to third order \cite{janik1,janik2}
\bea\label{Fexp}
F(w)/w\,=\, {2\over 3}+\, {1\over 3 w}\,\bar\eta \,-\,{1\over 3\,w^2} {\bar\eta\,(\ln 2- 1)\over 3\pi}\,+ 
\,{15-2\pi^2-45ln(2)+24(ln(2))^2 \over 972 \pi^3\,w^3}\,+ \, O(1/w^4)\,.
\eea  
The first term corresponds to the ideal hydro. The second one is NS, with $\bar\eta= 1/3\pi$, while the third one is the second order including non-linear terms, beyond IS.
At large $w$ the series is convergent.
 We will be arguing below that hydro is a reasonably good approximation for $w\ge w_0\simeq 0.4$. For illustration purpose we
give here  values of these terms at $w_0$, normalized to the first term:
\beq  (3/2)F(w_0)/w_0= 1+0.1326 +
                                0.0107
                               -0.0189 \,.
\eeq
 It is clear that the NS  term is still very important. The next terms are an order of magnitude smaller. Moreover, 
 we would like to stress the sign alternating feature of these  higher order terms.  As a result, being  resummed, these 
 terms  contribute less than each of them separately.
 
\begin{figure}[!h]
\begin{center}
\includegraphics[width=8cm]{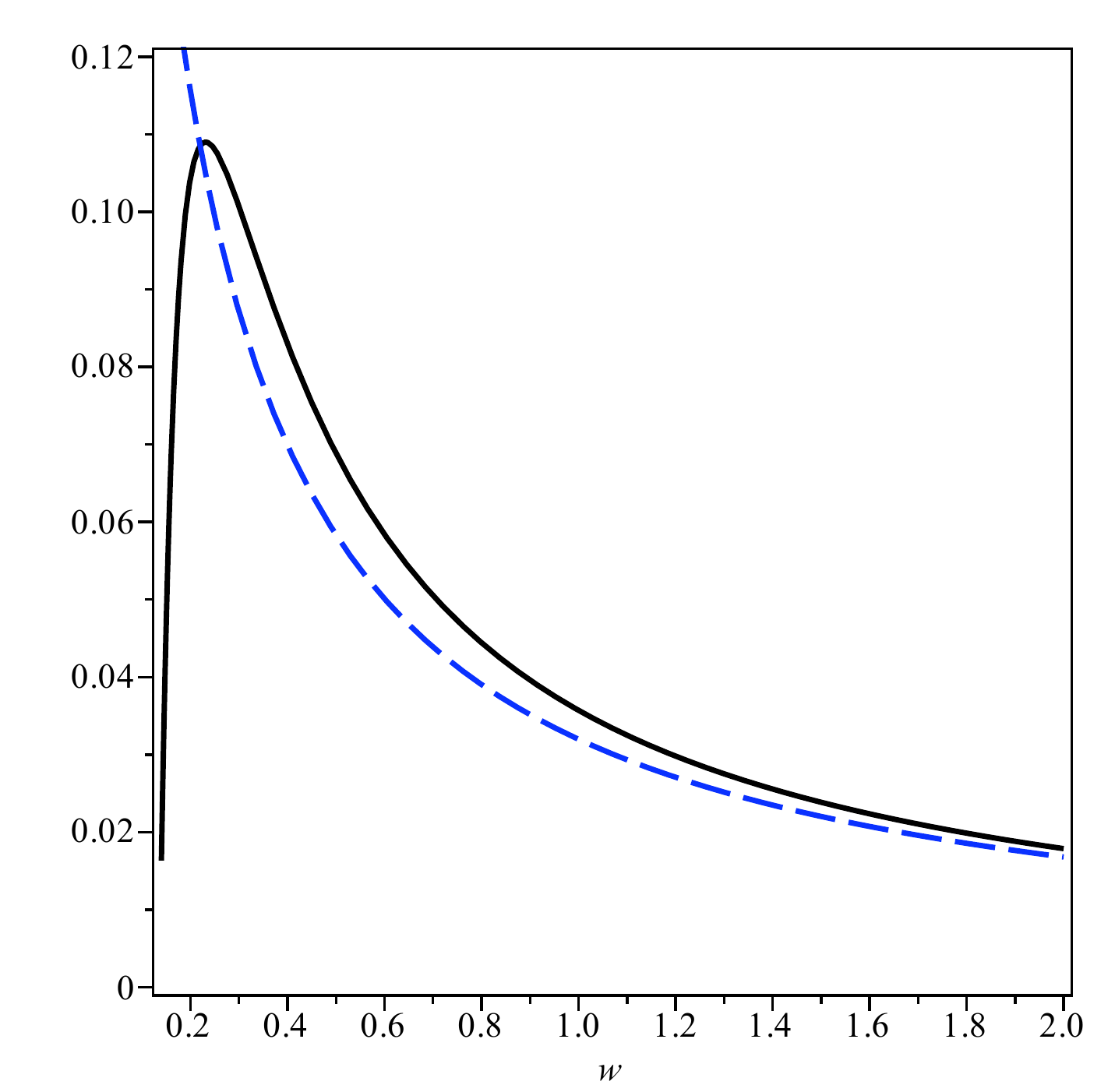}
\end{center}
%\vspace{-4cm}
\caption{(Color online) The solid line is  our model for  $F(w)/w-2/3$; the dashed line is its known large $w$ asymptotics (\ref{Fexp}).
}\label{modelfig}
\end{figure}

To get such qualitative behavior we proposed a new and very simple  ``resummation model" with a new (positive) parameter $\alpha$
\beq\label{model}
F(w)/w\,=\,  {2\over 3}+\, {\bar\eta \over 3\,(w+\alpha)}\,.
\eeq
This model obviously expands into a sign-alternating geometric series. The important feature is in the
 small $w$ behavior, which gets regularized.  One might want to relate $\alpha$  either to the relaxation time $\tau_R$ of IS or to the expansion terms in (\ref{Fexp}).
However, we are to argue  that the most natural choice is simply $\alpha=\bar\eta $: to eliminate any 
self heating at the early times, $\alpha$ shell be bigger than $\bar \eta$, $\alpha\ge\bar\eta$. $ \alpha=\bar\eta$ looks like the optimal model choice: it leads to $T(\tau)\sim\tau^0$ at small $\tau$ , which is consistent with \cite{Kovchegov} and CGC-based estimates. This choice
 maximizes the amount of entropy that can be produced within the model (\ref{model}). Larger $\alpha$ will drive the hydro to look  more ideal.
Fig.\ref{modelfig} compares this model function with the known asymptotics
at large $w$ given by (\ref{Fexp}).
%\vspace{-4cm}

% This would, however, lead 
%to negative $\alpha$, which is forbidden: to make sure the total entropy is not decreasing, $\alpha$ must be positive. Moreover, 

\subsection{AdS/CFT based studies of equilibration}
AdS/CFT correspondence provides a possibility to study strongly coupled  plasmas. We will not elaborate here on any details
but will only refer to some relevant results. 
Following  first applications  of the AdS/CFT to equilibrium properties (such as equation of state) and near-equilibrium kinetic coefficients (such as viscosity),
it was further realized that the duality provides a unique opportunity to study non-equilibrium problems  from  first principles based on
well-developed gravitational tools. From the 5th dimensional perspective, a fall of an object under  gravitational force  is equivalent to 
a relaxation process, proceeding from UV to IR. This was clearly demonstrated  in Ref. \cite{Lin} 
for an elastic membrane falling under its own weight .
%, from fully relaxed to unrelaxed scales.

Without citing the full list of the AdS/CFT-based studies of non-equilibrium phenomena, we would like to refer to two recent works \cite{Chesler:2010bi,janik2}, relevant for this note.  Both papers address the question to what extent an sQGP explosion 
 deduced from  exact numerical solutions of the Einstein equations in  AdS$_5$ agrees with a hydro evolution.
Relying on these studies one can answer the questions posed above, namely "what is hydro?" and "when does it start?", at least for the conformal plasma in study.
As seen from  Fig.3 of Ref. \cite{Chesler:2010bi}, full numerical solutions of the Einstein equations  agree with the NS hydro at quite early times. 
Similar analysis was performed in Ref.  \cite{janik2}. Starting  from a number of artificial initial conditions (which, to some extent, are equivalent to introducing  nuclei with
arbitrary structure functions) the authors of \cite{janik2} traced  the exact  time evolution from the gravity side.  It was found that, starting from 
some initial $w_i$, the evolution of {\bf all} trajectories converged to a universal behavior of the form (\ref{eq1}).  
%The purpose of it was to see $if$ and $when$ their evolution would be converged into an attractive fixed point, the hydrodynamics.
Fig.4 of this work displays this convergence and can be used to
%Furthermore, they do observe good convergence to some universal function $F(w)$, and thus can
define $w_i$ that  is  the ``beginning of hydro".  We conclude that, depending on the accuracy requested,
\beq   
w_i(few\,percents)\approx 0.40 \,\, \ \ \ \ \ \ \ \ \ \ \ \ w_i(half\,percent)\approx 0.65 \label{eqn_w_i}
\eeq
One of these values provides the second relation between $T_i$ and $\tau_i$, which, together with (\ref{sol}), fixes the initial conditions uniquely.
Obviously, our model function should be used for $w>w_i$ only. Its accuracy can be estimated
from comparison with the asymptotics (\ref{Fexp})  Fig.\ref{modelfig}.  As seen from the figure, the accuracy is about one percent or even better.
It is also important to note that in both studies  mentioned above, the convergence between the exact and hydro results happens 
 when the viscosity-induced asymmetries are still very large, $O(1)$. Emergence of  the ideal hydrodynamics (small asymmetries)
 can be also  seen in those results: it happens at noticibly later times.

\section{The entropy production}

The model (\ref{model}) makes it possible to consider a small $w$ limit with the function $F$ being well regularized.
Within this model, the proper time as a function of $w$ can be found analytically:
\beq
{\tau\over\tau_i}=\left ({w\over w_i}\right)^{3\alpha\over 2\alpha+\bar\eta}\,\left({2w+2\alpha+\bar\eta\over 2w_i+2\alpha+\bar\eta}\right)^{{3\over 2}-{3\alpha\over 2\alpha+\bar\eta}}
\eeq
%Note that $\alpha$ is negative, but pretty small, while $2\alpha+\bar\eta$ is positive.

The entropy density $s=4 k_B T^3$. Assuming  $R$, the ratio between the experimentally  measured multiplicity
and the pre-thermalization one, to coincide with the ratio between the finite and initial entropies,  we have
\beq
R={s\,\tau\over s_i\tau_i}= \left({w\over w_i}\ {2w_i+2\alpha+\bar\eta\over 2w+2\alpha+\bar\eta}\right)^{3-{6\alpha\over 2\alpha+\bar\eta}}\,.
\eeq
%Note that within the hydrodynamic approximation this ratio depends on the product $w_0$ only and not on $\tau_0$ and $T_0$ independently.
At the end of the evolution 
$$\tau_f\sim w_f^{3/2}\rightarrow\infty\,.
$$
$R$ goes to its limiting value 
$$
R=\left({2w_i+2\alpha+\bar\eta\over 2w_i} \right)^{3-{6\alpha\over 2\alpha+\bar\eta}}\,\simeq\, 1+{2\alpha+\bar\eta\over 2w_i} \left(3-{6\alpha\over 2\alpha+\bar\eta}\right)\,.
$$
For our choice $\alpha=\bar\eta$
\beq
R=\left({2w_i+3\bar\eta\over 2w_i} \right)  \approx 1.39\, ..\, 1.24\,,
\eeq
where the numerical values   $0.4\,..\,0.65$  were used for $w_i$.  Thus, our model  can nicely recover the missing 35\% in the total 
multiplicity production at the LHC at $E_{NN}=2.76 TeV$. This also supports $w_i\simeq 0.5$ as the right choice for the initial condition.

\subsection*{More on hydro initial conditions}

\begin{figure}[!h]
\begin{center}
\includegraphics[width=16cm]{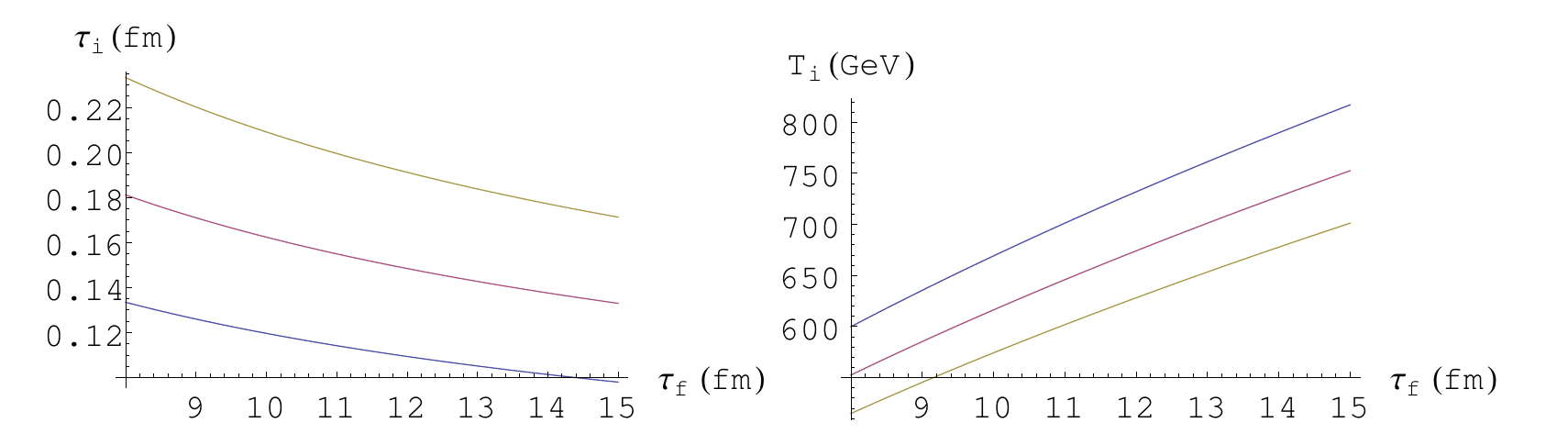}
\end{center}
\vspace{-5ex}\caption{ The initial time and temperature as a function of the freezout time $\tau_f$. Three curves correspond to $w_i=0.4\,(blue),\,0.5\,(red),\,0.6\,(yellow)$.
}\label{taui}
\end{figure}

As we argued above, hydro evolution (\ref{sol}), supplemented by a universal value of $w_i$ provides a means to estimate
both the initial temperature  $T_i$ and  initial time $\tau_i$ from the finite data. It makes sense to take as a final temperature $T_f$  the value of $170\,GeV$, 
being the QCD critical temperature.  The freezout time $\tau_f$ is not know well, neither we can be certain about our estimate of $w_i$.  Varying these parameters we can still provide a reasonable estimate for the initial data. We do it in Fig. \ref{taui}, which displays $\tau_i$ and $T_i$ as a function of $\tau_f$ for three values
of $w_i=0,4,\,0.5,\,0.6$.

\section{Summary and Discussion}

Inspired by the results derived via the AdS/CFT,  we argued that for a rapidity-independent geometry, a non-equilibrium explosion
rapidly converges  to a universal hydrodynamical function $F(w)$, for which  we proposed a new simple model. Using this model we
estimated the amount of entropy produced in the hydrodynamic stage and found it to constitute about 30\% of the total. This
 compliments  perturbative studies such as of Ref. \cite{Albacete}, and recovers  the ``missing entropy" in heavy ion collisions at the LHC.
 In addition, we were able to provide estimates both for the initial temperature $T_i\sim 600-800\, GeV$ and for the thermalization time
 $\tau_i\sim 0.1-0.2\,fm$.

The key question, of course, is to what extent  our arguments and estimates are indeed applicable to QCD  in real heavy ion collisions.
QCD plasma is believed to be strongly coupled.  From the lattice studies,  thermodynamics of QCD is also known to display a  near-conformal behavior
 ($p/T^4,\epsilon/T^4\sim const$ for $T>2T_c$).  However, QCD is presumably not a theory with a gravity dual, and there is little information about its 
 transport coefficients.
% Yet we do not know 
 %the values of the whole series of viscosities.
In particular,  there is no reason for the  viscosity to entropy  ratio to have the same universal value $1/4\pi$ as in theories with gravity duals. 
% $\bar\eta$ should have the universal KSS  \cite{KSS} value.
Indeed, presently available phenomenological estimates of the viscosity favor a larger value for the ratio: Ref.  \cite{Shen:2011zc}
obtained  $\eta/s\simeq2\ (1/4\,\pi)$ at the RHIC and $\eta/s\simeq2.5\ (1/4\,\pi)$ at the LHC. First studies of the higher flow harmonics 
support these findings \cite{Lacey:2011av,Staig:2011wj}.
 Furthermore,  this ratio is expected to increase  with the temperature, because the coupling  becomes weaker.
Perturbative studies, such as of \cite{Xu:2007jv}, relate $\eta/s$ to an interplay between $gg\rightarrow gg$ and $gg\rightarrow ggg$ cross sections, which are
$O(\alpha_s(T)^2)$ and $O(\alpha_s(T)^3)$ respectively (processes with more gluons in the final state can be also considered \cite{Xiong:1992cu}). 
%It is hard to get the absolute numbers, though, as the exact coupling running so close to $T_c$ are not yet well determined. 

Non-perturbative studies relate viscosity  to an interplay between gluon-gluon  and gluon-monopole scattering processes. 
In the latter case, there is no coupling constant in the cross section (electric charge times magnetic  is integer); 
the  monopole density, computed on the lattice, decreases with temperature. 
Ref. \cite{Ratti:2008jz}  predicted a  rise in  $\eta/s$ as a function of temperature (Fig. 14 of this paper),
 from $\eta/s\simeq2\ (1/4\,\pi)$ at $T=2T_c$ to $\eta/s\simeq2.6\ (1/4\,\pi)$ at $T=4T_c$, roughly corresponding to the initial conditions at the RHIC and LHC.
 % initial temperatures: so \cite{Ratti:2008jz}  predicts about 30\% growth of the viscosity ratio from RHIC to LHC.

If the QCD plasma were conformal, $R$ would not depend on the collision energy $E_{NN}$, and we would obtain the same 
prediction for the RHIC and LHC.  However, as we have noted in the beginning, experimentally it is not true. 
We are to speculate  that this extra multiplicity  observed  at the LHC (relative to the RHIC normalization) 
  may originate from the viscosity growth as a function of temperature.
We further conjecture that  our ``universal resummed hydrodynamics" should, in some form, be valid in any theory
and  perhaps the same value of $w_i$ parameter will is true in QCD.
Then, the extra entropy produced, between the RHIC and LHC, can be ascribed  to  viscosity growth. 
Relying on  our ``resummed hydrodynamics" result we get
\beq
{ R(LHC)\over R(RHIC)}\,\approx 1+{3[\bar\eta(LHC)-\bar\eta(RHIC)]\over  2w_i+3\bar\eta(RHIC)}  
\eeq
Substituting  $\bar\eta(RHIC)\approx 2\, (1/3\pi)$,
$\omega_i=0.4$ we find that in order to get the $23-5.5=16.5$\% (\ref{dr}) of the 
unaccounted extra multiplicity (double ratio) growth at LHC
 one would need  the relative viscosity growth $[\bar\eta(LHC)-\bar\eta(RHIC)]/\bar\eta(RHIC) \simeq 0.4$,
 which is in the expected ballpark. 

The ultimate knowledge about QCD transport properties will come from a systematic study of various hydrodynamical phenomena,
beyond entropy production discussed in this note. The most promising ones are sound waves,  already discussed in the Introduction.

%Now, how can one further test the proposed explanation? By looking at hydrodynamical phenomena other than
%entropy production. The initial state fluctuations create sound waves, which are observed in the correlation functions, right now
%up to the 9-th angular harmonics, both at RHIC and LHC (see e.g. \cite{Lacey:2011av,Staig:2011wj} and references therein). 
%The ``power spectrum" of those harmonics will produce more definite  estimates of the QGP viscosity soon, based on sound amplitudes
%which are completely
%independent from the multiplicities.  

%Summarizing: We have shown that about 30\% of the entropy observed in AA collisions come from the hydro stage. 
%This explains a discrepancy with perturbative calculations. We then speculated further, that
 %extra multiplicity  observed  at the LHC (relative to RHIC normalization) 
  %may originate from the viscosity growth. (The latter happens because the initial temperature at LHC is higher than at RHIC
  %and the QGP is somewhat weaker coupled there.) 

\section*{Acknowledgments}
ML is very grateful to the BNL and Stony Brook Nuclear Theory Groups for the hospitality  during the period when this work was completed.
The work of ML is partially supported by the Marie Curie Grant  PIRG-GA-2009-256313.

\end{document}